\documentclass[aps,prl,twocolumn,superscriptaddress,notitlepage]{revtex4-1}

\usepackage{graphicx}
\usepackage{amsmath}
\usepackage{amssymb}
\usepackage{mathtools}
\usepackage{multirow}
\usepackage{makecell}
\usepackage{multirow}
\usepackage{color}

\bibpunct{[}{]}{;}{n}{}{}

\DeclareGraphicsExtensions{.png,.jpg,.eps}
\usepackage{xcolor}

\def\bh{{\bold{h}}}
\def\bk{{\bold{k}}}

\def\m1{{^{-1}}}

\begin{document}


\title{Two scenarios for superconductivity in CeRh$_2$As$_2$}

\author{David  M\"{o}ckli}

\affiliation{Instituto de F\'{i}sica, Universidade Federal do Rio Grande do Sul, 91501-970 Porto Alegre, RS, Brazil}

\author{Aline Ramires}
\affiliation{Paul Scherrer Institut, CH-5232 Villigen PSI, Switzerland}

\begin{abstract}
CeRh$_2$As$_2$, a non-symmorphic heavy fermion material, was recently reported to host a remarkable phase diagram with two superconducting phases. In this material, the two inequivalent Ce sites per unit cell, related by inversion symmetry, introduce a sublattice structure corresponding to an extra internal degree of freedom. Here we propose a classification of the possible superconducting states in CeRh$_2$As$_2$ from the two Ce-sites perspective. Based on the superconducting fitness analysis and the quasiclassical Eilenberger equations, we discuss two limits: Rashba spin-orbit coupling and inter-layer hopping dominated normal state. In both limits, we are able find two scenarios that generate phase diagrams in qualitative agreement with experiments: i) intra-sublattice pairing with an even-odd transition under magnetic field, and ii) inter-sublattice pairing with an odd-odd transition under magnetic field.
\end{abstract}

\date{\today}

\maketitle


The heavy fermion CeRh$_2$As$_2$ was recently discovered to host remarkable properties in the  superconducting state \cite{Khim:2021}. Under c-axis magnetic field, a transition between a low-field and a high-field superconducting phase is observed by measurements of magnetization and magnetostriction. The presence of two superconducting phases is unusual, and has only been reported in other two stoichiometric heavy fermion materials: UPt$_3$ \cite{Fischer:1989,Adenwalla:1990,Joynt:2002} and  UTe$_2$ \cite{Ran:2019}. The high effective electronic mass inferred from the low temperature value of the specific heat coefficient indicates that CeRh$_2$As$_2$ is in the heavy fermion regime, but power-law temperature dependence of the specific heat below 4K suggests the proximity to a quantum critical point, whose fluctuations can be playing an important role for pairing \cite{Dyke:2014,Landaeta:2018,White:2015,Smidman:2018}. Furthermore, the high-field phase has an upper critical field of 14T, exceptional for a material with a superconducting critical temperature ($T_c$) of 0.26K, suggesting that CeRh$_2$As$_2$ hosts an unconventional triplet superconducting state.

Theory developed in the context of layered superconductors with local inversion symmetry breaking accounts for the qualitative features of the temperature versus magnetic field phase diagram of CeRh$_2$As$_2$ \cite{Maruyama:2012,Yoshida:2014}. More recently, these studies were supplemented by detailed investigations including also orbital depairing effects \cite{Mockli:2018a,Schertenleib:2021}. Intra-layer singlet pairing (referred here as BCS) is assumed to be the stable superconducting state within each layer. Once a magnetic field is applied perpendicular to the layers, a pair-density wave (PDW) state, with a sign change of the order parameter between layers, is favoured under the requirement that Rashba spin orbit coupling (SOC) is comparable to inter-layer hopping (ILH) amplitudes.

Nevertheless, the phenomenology of CeRh$_2$As$_2$ indicates that other scenarios might be possible. In particular, the small anisotropy of the effective mass, inferred from the slope of the upper critical field around $T_c$, indicates that the system is rather three-dimensional \cite{Khim:2021}. This is in agreement with recent first-principles calculations \cite{Ptok:2021}, and with other 122-materials in this family. As an example, CeCu$_2$Si$_2$ crystalizes in the ThCr$_2$Si$_2$-type structure (the centrosymmetric analog of the CaBa$_2$Ge$_2$-type structure of CeRh$_2$As$_2$), displays a 3-dimensional spin density wave state supported by Fermi surface nesting, and  hosts superconductivity around the pressure induced quantum critical point \cite{Arndt:2011,Steglich:2012}.

\begin{center}
\begin{figure}[t]
\includegraphics[width=0.8\linewidth, keepaspectratio]{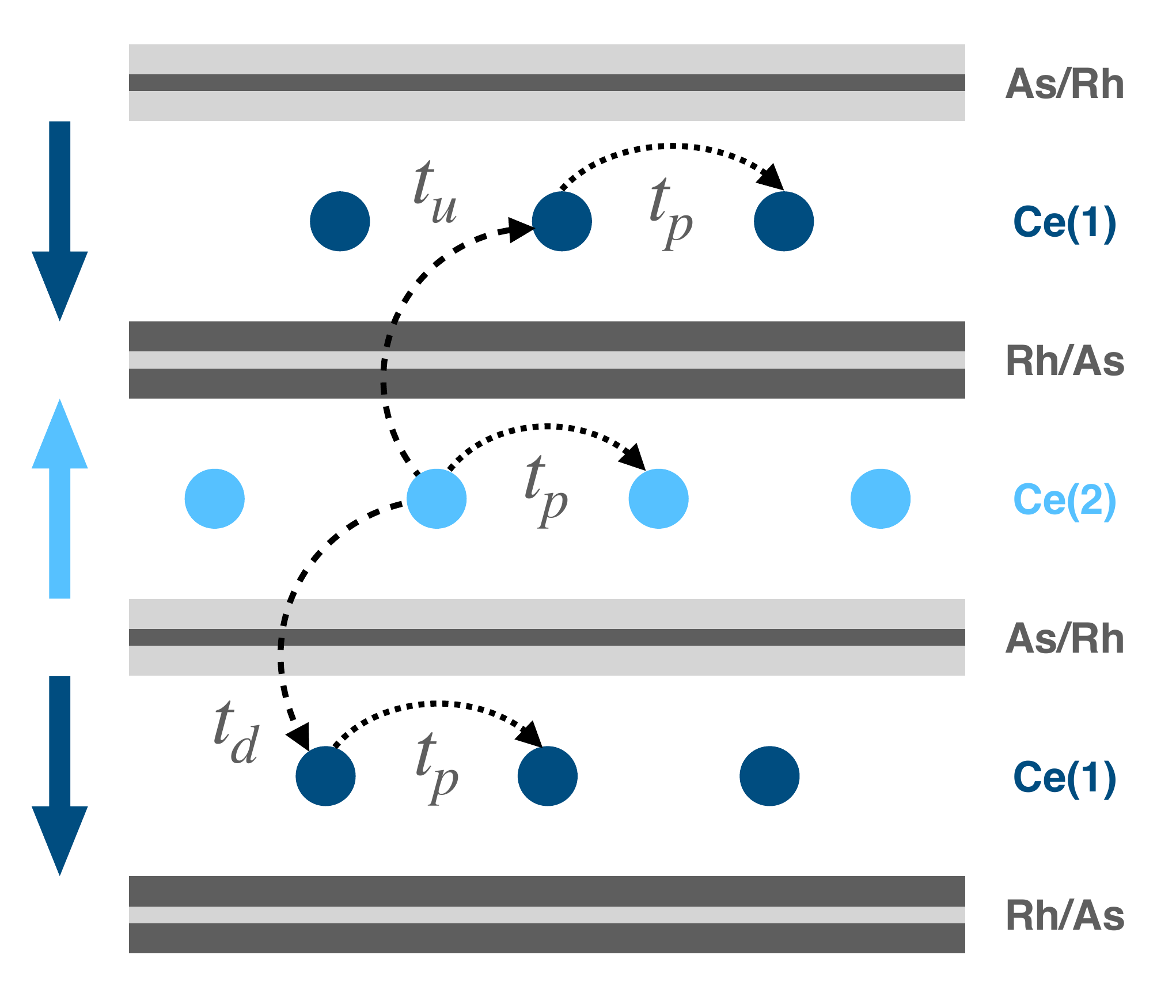}
\caption{ Schematic depiction of CeRh$_2$As$_2$. The spheres correspond to the two inequivalent Ce atoms and the rectangular structures to the two types of Rh/As layers. The big arrows on the left correspond to the effective electrical field generating a staggered Rashba-SOC. The dotted lines correspond to in-plane intra-SL hopping with amplitude $t_p$ (same for all layers), and the dashed lines correspond to inter-SL hopping, with amplitude $t_u$ or $t_d$, depending if the hopping is to a neighbor in the layer above or below. 
}
\label{Fig:Model}
\end{figure}
\end{center}

Motivated by these facts, here we analyse the possible unconventional superconducting states that can be hosted by CeRh$_2$As$_2$. We start with focus on the main ingredient, the Ce ions, and propose a microscopic model for the normal state Hamiltonian based on the spin and sublattice (SL) degrees of freedom (DOF). This model allows us to discuss two limits: 
i) two-dimensional, dominated by Rashba-SOC, and ii) three-dimensional, dominated by ILH. We then classify all possible types of Cooper pairs that can be formed in this material. We perform the superconducting fitness analysis and discuss the relative stability of different superconducting states in both limits. Finally, we obtain temperature versus magnetic field phase diagrams from the quasiclassical Eilenberger equations, based on which we can propose two scenarios for superconductivity in CeRh$_2$As$_2$.

CeRh$_2$As$_2$ has a CaBa$_2$Ge$_2$-type structure with the non-symmorphic  centrosymmetric space group \emph{P4/nmm} (No. 129).  Given the heavy fermion nature of the electronic structure around the Fermi energy inferred by recent experiments, we start modelling the electronic DOF from the Ce sites perspective. The Ce atoms are located in between Rh-As layers which appear intercalated in two flavors: with Rh atoms tetrahedrally coordinated by As, or vice-versa, as schematically shown in Fig. \ref{Fig:Model}. The intercalation of two types of Rh-As layers generates two inequivalent Ce sites with $C_{4v}$ point group symmetry. Importantly, the Ce sites are not centers of inversion.  The point group $C_{4v}$ can be generated by $C_{4z}$, a rotation along the z-axis by $\pi/2$, and $\sigma_{xz}$, a mirror reflection along the xz-plane. The complete space group includes inversion, which can be made a symmetry at the Ce sites if composed with half-integer translation vectors. We define $i_{1/2}$ as inversion composed with a translation by $(a/2,a/2,c/2)$, which links the two inequivalent Ce sites (here $a$ and $c$ are the unit cell dimensions in the plane and along the z-axis, respectively). These three operations generate the complete space group. Here we note that this space group is isomorphic to $D_{4h}$ up to integer lattice translations  \cite{Cvetkovic:2013}. Given these generators, Table \ref{Tab:Irreps} summarizes the properties of the irreducible representations. For simplicity, here we focus on the symmetry analysis around the $\Gamma$ point.

We start with the most general non-interacting normal state Hamiltonian considering Wannier functions localised at the Ce atoms accounting for a two SL structure:
\begin{eqnarray}
\mathcal{H}_0= \sum_\bk\hat{\Psi}^\dagger_\bk \hat{H}_0(\bk) \hat{\Psi}_\bk,
\end{eqnarray}
where $\hat{\Psi}^\dagger=(c_{1\uparrow}^\dagger,c_{1\downarrow}^\dagger,c_{2\uparrow}^\dagger,c_{2\downarrow}^\dagger )$ encodes the two SLs $(1,2)$ and the spin $(\uparrow, \downarrow)$  DOF. The $4\times 4$ matrix $\hat{H}_0(\bk)$ can be parametrized as:
\begin{eqnarray}\label{Eq:H0}
\hat{H}_0 (\bk)= \sum_{a,b} h_{ab}(\bk) \hat{\tau}_a\otimes\hat{\sigma}_b,
\end{eqnarray}
where $\hat{\tau}_a$ and $\hat{\sigma}_b$ are Pauli matrices, $\{a,b\} = \{1,2,3\}$,  or the two-dimensional identity matrix, $\{a,b\} = \{0\}$, corresponding to the SL and spin DOF, respectively. 
In presence of inversion (implemented as $\tau_1\otimes \sigma_0$ accompanied by $\bk \rightarrow -\bk$) and time-reversal symmetry (implemented as $i \tau_0 \otimes \sigma_2$, accompanied by complex conjugation and $\bk \rightarrow -\bk$), only the $(a,b)$ pairs summarized in Table \ref{Tab:H0} are symmetry allowed. 
 
 \begin{table}[t]
\begin{center}
    \begin{tabular}{| c | c | c | c | c |}
    \hline
Irrep & $C_{4z}$ & $\sigma_{xz}$ & $i_{1/2}$ & Basis\\ \hline
$A_{1g}$ & +1 & +1 & +1 & $x^2+y^2$, $z^2$ \\ \hline
$A_{2g}$ & +1 & -1 & +1 & $xy(x^2-y^2)$ \\ \hline
$B_{1g}$ & -1 & +1 & +1 & $x^2-y^2$ \\ \hline
$B_{2g}$ & -1 & -1 & +1 & $xy$ \\ \hline
$E_{g}$ & 0 & 0 & +2 & $\{xz, yz\}$ \\ \hline
$A_{1u}$ & +1 & +1 & -1 & $z$ \\ \hline
$A_{2u}$ & +1 & -1 & -1 & $xyz(x^2-y^2)$ \\ \hline
$B_{1u}$ & -1 & +1 & -1 & $z(x^2-y^2)$ \\ \hline
$B_{2u}$ & -1 & -1 & -1 & $xyz$  \\ \hline
$E_{u}$ & 0 & 0 & -2 & $\{x, y\}$ \\ \hline
    \end{tabular}
        \end{center}
    \caption{  \label{Tab:Irreps}  Irreducible representations (irrep) at the $\Gamma$-point associated with the three symmetry operations at the Ce sites. The last column shows examples of polynomials in each irrep.}
\end{table}

\begin{table}[t]
\begin{center}
    \begin{tabular}{| c | c | c | c | c | }
    \hline
    $(a,b)$ & Irrep & $\bk$&  Process & Parameter    \\ \hline
    $(0,0)$ & $ A_{1g}$ &  Even  & Intra-SL hopping & $t_p$   \\ \hline
    $(1,0)$ & $ A_{1g}$ & Even  & Inter-SL hopping  & $t_u+t_d$ \\ \hline
    $(2,0)$ & $ A_{1u}$ &  Odd   & Inter-SL hopping  & $t_u-t_d$ \\ \hline
    $(3,1)$ &  \multirow{2}{*}{ $E_{u}$} & \multirow{2}{*}{ Odd }  & \multirow{2}{*}{ Rashba-SOC} &  \multirow{2}{*}{$\alpha$ }   \\ \cline{1-1}
    $(3,2)$ & &  & &  \\ \hline
    $(3,3)$ & $ A_{2u}$ & Odd   & Ising-SOC & $\lambda$  \\ \hline
    \end{tabular}
        \end{center}
    \caption{  \label{Tab:H0} Symmetry allowed $(a,b)$ terms in the normal state Hamiltonian, as given in Eq. \ref{Eq:H0}. The table highlights the irrep, the even or odd $\bk$ dependence, the associated physical process and dominant parameter for each term.}
\end{table}

We now associate each term with specific physical processes. $(0,0)$ concerns intra-SL hopping, with $h_{00}(\bk) = 2 t_p [\cos (k_xa) + \cos (k_ya)]$, dominated by intra-layer hopping to nearest neighbours with amplitude $t_p$. $(1,0)$ and $(2,0)$ stem from inter-SL hopping which in this case is necessarily out-of-plane, with $h_{10}(\bk) = 4 (t_u + t_d) \cos (k_xa/2)\cos (k_ya/2) \cos (k_zc/2)$ and $h_{20}(\bk) = -4(t_u - t_d) \cos (k_xa/2)\cos (k_ya/2) \sin (k_zc/2)$. These terms are parametrised by $t_u$ and $t_d$, the hopping amplitudes to the nearest neighbors in the layer above and below, which are inevitably distinct due to inversion symmetry breaking. $(3,1)$ and $(3,2)$ are concerned with intra-SL staggered Rashba-type SOC, $h_{31}(\bk) = -\alpha \sin (k_y a)$ and $h_{32}(\bk) = \alpha \sin (k_x a)$, parametrized by $\alpha$. Finally, $(3,3)$ is an Ising-type SOC, $h_{33}(\bk) = \lambda \sin (k_x a) \sin (k_y a) \sin (k_z c)[\cos (k_x a) - \cos (k_y a)]$, associated with hopping to neighbors of the same SL in the next nearest layer, parametrized by $\lambda$. This Hamiltonian is the same as the one proposed in Ref. \cite{Khim:2021}. 


The superconducting order parameter can be written in a similar fashion:
\begin{eqnarray}\label{Eq:D}
\hat{\Delta}(\bk)= \sum_{a,b} d_{ab}(\bk) \hat{\tau}_a\otimes\hat{\sigma}_b (i\sigma_2).
\end{eqnarray}
In presence of inversion symmetry, we can distinguish between even or odd in $\bk$ and even or odd parity gap functions. Table \ref{Tab:SCAll}  lists all the order parameters, their nature in terms of the microscopic DOFs, and their symmetry properties. The different brackets distinguish the normal state Hamiltonian parameters, $(a,b)$, from the superconducting order parameters, $[a,b]$.

\begin{table}[t]
\begin{center}
    \begin{tabular}{| c | c | c | c | c | c |  }
    \hline
     Irrep&  $[a,b]$  & Spin & SL & $\bk$ & Parity  \\ \hline
\multirow{2}{*}{ $A_{1g}$ }& $[0,0]$ &  S  & Intra & E & E \\\cline{2-6}
& $[1,0]$ &   S  & Inter & E & E  \\ \hline
\multirow{2}{*}{ $A_{2g}$ }& $[0,3]$ &  T  & Intra & O & O  \\ \cline{2-6}
& $[1,3]$  &  T  & Inter & O & O  \\ \hline
\multirow{2}{*}{ $A_{1u}$ }& $[3,0]$ &  S  & Intra & E & O    \\ \cline{2-6}
& $[2,0]$ &   S  & Inter & O & E  \\ \hline
\multirow{2}{*}{ $A_{2u}$ }& $[2,3]$ &   T   & Inter & E & O  \\ \cline{2-6}
& $[3,3]$ &   T  & Intra & O & E  \\ \hline
\multirow{2}{*}{ $E_g$ }& $\{[0,1],[0,2]\}$ &   T  & Intra & O & O  \\ \cline{2-6}
 & $\{[1,1],[1,2]\}$ & T  & Inter & O & O \\ \hline
\multirow{2}{*}{ $E_u$ }&$\{[2,1],[2.2]\}$ & T   & Inter & E & O   \\ \cline{2-6}
&$\{[3,1],[3,2]\}$ &  T  & Intra & O & E  \\ \hline
    \end{tabular}
        \end{center}
    \caption{  \label{Tab:SCAll} Symmetry classification of the $[a,b]$ matrices associated with all order parameters, defined in Eq. \ref{Eq:D}, organized by irreducible representations (irreps) around the $\Gamma$ point. Here E/O stands for even/odd and S/T for singlet/triplet. The irrep associated with the complete order parameter is obtained by taking the product with the irrep of $d_{ab}(\bk)$, which is always nontrivial for the odd-$\bk$ order parameters.}
\end{table}

Within the standard weak-coupling formalism, the superconducting fitness analysis allows us to discuss the relative stability of  superconducting states based on properties of the normal electronic state \cite{Ramires:2016,Ramires:2018}. In particular, it was shown that the larger the average over the FSs of $\text{Tr}\,[\hat{F}_A(\bk)^\dagger \hat{F}_A(\bk)]$, the more robust the superconducting instability and the higher critical temperature. Here, $\hat{F}_A(\bk) =\tilde{ H}_0(\bk) \tilde{\Delta}(\bk) + \tilde{\Delta}(\bk) \tilde{H}_0^*(-\bk)$ is the superconducting fitness matrix, written in terms of the normalized normal state Hamiltonian $\tilde{H}_0(\bk) = [\hat{H}_0(\bk) - h_{00}(\bk)\tau_0\otimes \sigma_0]/|\bh(\bk)|$, where $|\bh(\bk)| = \sqrt{\sum_{(a,b)\neq (0,0)}|h_{ab}(\bk)|^2}$,  and the normalized gap matrix $\tilde{\Delta}(\bk)$  with $\langle\tilde{\Delta}(\bk) \rangle_{FS} =1$.

The superconducting fitness analysis for CeRh$_2$As$_2$ is summarized in Table \ref{Tab:FA}. We start discussing the scenario with dominant ILH, such that we assume $|t_p|> |t_u+t_d|> |t_u-t_d| > |\alpha| >  |\lambda|$. From Table \ref{Tab:FA}, we find that Cooper pairs with $a=0$ are the most stable since they have the contribution from the two largest terms in the normal state Hamiltonian, $(1,0)$ and $(2,0)$. These are followed by pairs with $a=1$, supported only by $(1,0)$ hopping, controlled by $(t_u+t_d)^2$. Pairs with $a=2$ are less stable since these are supported only by $(2,0)$, controlled by $(t_u-t_d)^2$, while pairs with $a=3$ are not stabilized by any hopping term in the normal state Hamiltonian.  Among the intra-SL pairs $(a=0,3)$, the most robust order parameters are the spin singlets, stabilized by the larger contribution stemming from both Rashba and Ising SOC. Among the inter-SL pairs $(a=1,2)$, the most stable order parameters are the spin triplet states with a d-vector along the z-direction, stabilized by Rashba SOC. Overall this analysis supports $[0,0]$ and $[1,3]$ as the two most robust superconducting states for the ILH dominated scenario. Moving now to the SOC dominated scenario, we assume $|t_p|> |\alpha| > |t_u+t_d|> |t_u-t_d| > |\lambda|$. For the intra-SL Cooper pairs ($a=0,3$), we find that SOC stabilizes spin singlet states ($b=0$), while for inter-SL pairs SOC stabilizes triplet states with d-vector along the z-axis ($b=3$). Among the intra-SL pairs, a finite ILH stabilizes $[0,0]$, and among the inter-SL pairs it supports $[1,3]$. Note that these are the same order parameter candidates found for the ILH dominated scenario.

\begin{table}[t]
\begin{center}
    \begin{tabular}{| c | c | c | c | c | c  | c | c |}
    \hline
 & (1,0) & (2,0) & (3,1) & (3,2) & (3,3) & ILH & SOC\\ \hline
$[0,0]$  & 1 & 1 & 1 & 1 & 1 & \multirow{4}{*}{$2(t_u^2+t_d^2)$} & $2\alpha^2+\lambda^2$  \\ \cline{1-6}\cline{8-8}
$[0,1]$  & 1 & 1 & 1 & 0 &0 & & \multirow{2}{*}{$\alpha^2$} \\ \cline{1-6}
$[0,2]$  & 1 & 1 & 0 & 1 & 0 & &  \\ \cline{1-6}\cline{8-8}
$[0,3]$  & 1 & 1 & 0 & 0 & 1 & & $\lambda^2$ \\ \hline
$[1,0]$  & 1 & 0 & 0 & 0 & 0 & \multirow{4}{*}{$(t_u+t_d)^2$}& 0 \\ \cline{1-6}\cline{8-8}
$[1,1]$  & 1 & 0 & 0 & 1 & 1 & &  \multirow{2}{*}{$\alpha^2 +\lambda^2$} \\ \cline{1-6}
$[1,2]$  & 1 & 0 & 1 & 0 & 1 & &  \\ \cline{1-6}\cline{8-8}
$[1,3]$  & 1 & 0 & 1 & 1 & 0 & & $2\alpha^2$  \\ \hline
$[2,0]$  & 0 & 1 & 0 & 0 & 0 &  \multirow{4}{*}{$(t_u-t_d)^2$}& 0 \\ \cline{1-6}\cline{8-8}
$[2,1]$  & 0 & 1 & 0 & 1 & 1 &  &  \multirow{2}{*}{$\alpha^2 +\lambda^2$} \\ \cline{1-6}
$[2,2]$  & 0 & 1 & 1 & 0 & 1 &  &  \\ \cline{1-6}\cline{8-8}
$[2,3]$  & 0 & 1 & 1 & 1 & 0 &  &  $2\alpha^2$ \\ \hline
$[3,0]$  & 0 & 0 & 1 & 1 & 1 & \multirow{4}{*}{0} & $ 2\alpha^2+\lambda^2$\\ \cline{1-6}\cline{8-8}
$[3,1]$  & 0 & 0 & 1 & 0 & 0 &  &  \multirow{2}{*}{$\alpha^2$} \\ \cline{1-6}
$[3,2]$  & 0 & 0 & 0 & 1 & 0 &  &  \\ \cline{1-6}\cline{8-8}
$[3,3]$  & 0 & 0 & 0 & 0 & 1 &  & $ \lambda^2$ \\ \hline
    \end{tabular}
        \end{center}
    \caption{  \label{Tab:FA} Superconducting fitness analysis for all order parameters. Each line corresponds to an order parameter labelled by $[a,b]$, as in Eq. \ref{Eq:D}. The numerical entries correspond to $\text{Tr}\,[\hat F_{A}(\bk,s)^\dagger \hat F_{A}(\bk,s)] = 4 \sum_{cd} (\text{table entry})\, |h_{cd}(\bk,s)|^2/|h(\bk,s)|^2$, for each term $(c,d)$ in the normal-state Hamiltonian. The last two columns summarize the effects of ILH and SOC dropping the accompanying momentum dependence of the respective functions.}
\end{table}

Within the rough assumption that pairing in all channels has same strength, we can conclude that the trivial order parameter with matrix structure $[0,0]$, an even parity intra-SL spin singlet (BCS) state, is the absolutely most stable superconducting  state since all terms in the normal state Hamiltonian support the instability in this channel. We note here that within the assumption of $t_u=t_d$ and $\lambda=0$, the $[1,3]$ order parameter candidate, an odd parity SL-symmetric spin triplet (SLS-ST), would have the same transition temperature. These are good candidates for the low-field superconducting phase of CeRh$_2$As$_2$.

In presence of a magnetic field along the z-axis, the proposed PDW state, an odd parity intra-SL spin singlet, here captured by $[3,0]$, is not very robust since the normal state ILH terms $(1,0)$ and $(2,0)$ do not contribute to its stabilisation. Interestingly, from the analysis above, within the assumption of $t_u=t_d$ and $\lambda=0$, the order parameter $[2,3]$, an odd parity SL-antisymmetric spin triplet (SLA-ST), is equally stable. In fact, under the assumption $|t_u-t_d|>\lambda$, the SLA-ST is more robust than the PDW state. Order parameters with a similar nature, antisymmetric in an internal DOF, were recently proposed in multiple contexts \cite{Suh:2020,Ghosh:2020,Vafek:2017,Fu:2010}.

\begin{center}
\begin{figure}[t]
\includegraphics[width=\linewidth, keepaspectratio]{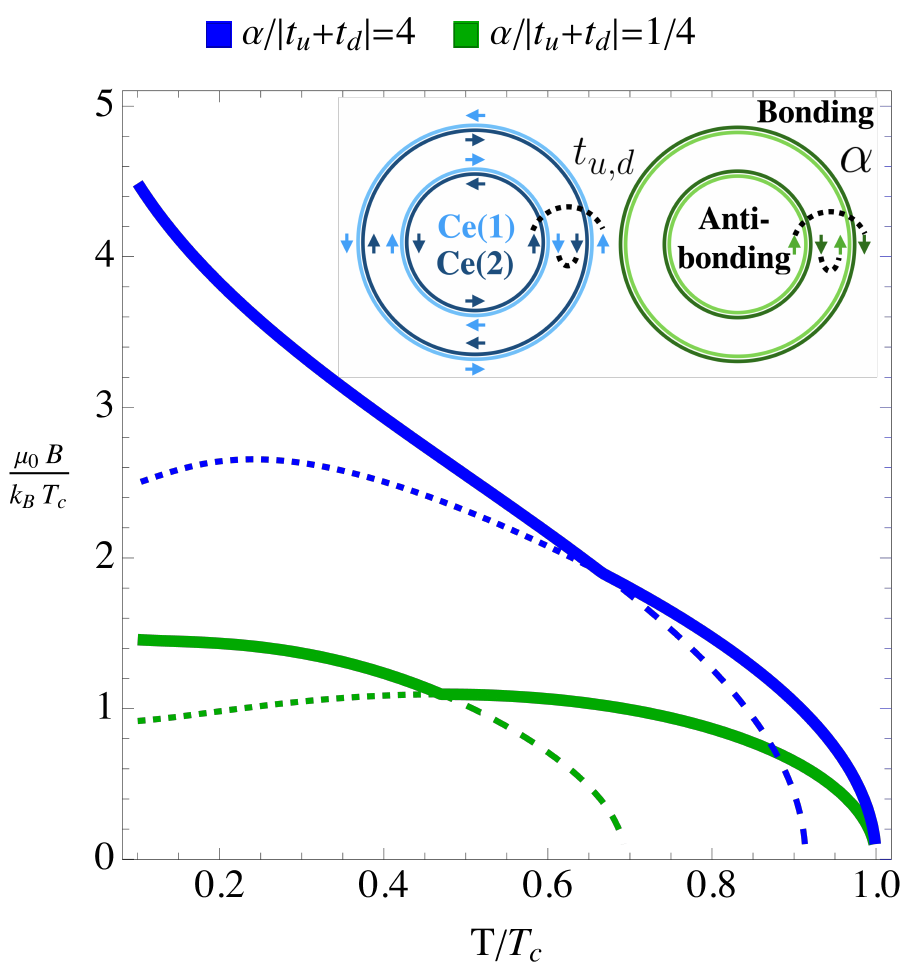}
\caption{
Identical phase diagrams for the even-odd intra-SL $[0,0] \rightarrow [3,0]$ and odd-odd inter-SL $[1,3]\rightarrow[2,3]$ scenarios with $|t_u+t_d|=T_c$. The blue curves correspond to the SOC-dominated regime, while the green curves correspond to the ILH-dominated regime. The dotted (dashed) lines are the extensions of the low (high) field solutions. 
The inset displays schematic two-dimensional cuts of the FSs. Left: SOC-dominated regime. The lighter (darker) color indicates FSs stemming from Ce(1) [Ce(2)] layers, and the arrows indicate the helicity. Turning on ILH connects FSs with same helicity and complementary Ce-content (dotted lines). Right: ILH-dominated regime. The lighter (darker) color indicates spin up (down). Turning on SOC connects FSs with opposite spin and bonding state (dotted lines).
}
\label{Fig:PD}
\end{figure}
\end{center}

With these facts at hand, we now examine the behaviour under magnetic field of the four order parameters identified above.
The hierarchy of energy scales motivates the writing of a quasiclassical theory assuming $|t_p| \gg  \{|t_u+t_d|,|t_u-t_d|,\alpha,\lambda\}$. For simplicity, here we take $t_u=t_d$ and $\lambda=0$ and study the interplay of the magnetic field $B$, Rashba SOC $\alpha$ and ILH $|t_u+t_d|$. We extend the linearized Eilenberger equations of Ref. \cite{Mockli:2020} to include $|t_u+t_d|$, which allows us to obtain the transition lines of the best low and high field phase candidates. For each pairing irrep-channel, we associate a superconducting critical temperature $T_\mathrm{irrep}$ that is defined in the absence of magnetic field and SOC. For simplicity, we consider the same critical temperatures for all fitness favored channels by setting $T_{A_{1g}}=T_{A_{2g}}=T_{A_{1u}}=T_{A_{2u}}=T_c$.
We find two promising scenarios that display the same phase diagram: (i) inter-SL scenario, with a transition from a low field odd SLS-ST state $[1,3]$ to a high field odd SLA-ST state $[2,3]$; (ii) intra-SL scenario, with a transition from an even BCS state $[0,0]$ to a high field odd PDW state $[3,0]$.
These results are summarized in Fig. \ref{Fig:PD}. In both scenarios, a first-order phase transition separates the high from the low field phase. The exact location of these transitions requires a treatment beyond linearization.


Given the extremely large upper critical field observed experimentally, the discussion above suggests that the normal state is SOC dominated. Note, though, that this limit does not allows us to distinguish between the intra-SL and inter-SL scenarios, as illustrated by Fig. \ref{Fig:PD}. One potential way to distinguish them is the effect of impurities. Current samples of CeRh$_2$As$_2$ are likely to be in the dirty limit \cite{Khim:2021}, in which case a $\bk$-independent order parameter in the microscopic basis would be robust \cite{Mockli:2018b,Mockli:2020,MockliJAP:2020,Andersen:2020}. This suggests the intra-SL scenario for the low field phase, since in the inter-SL $d_{13}(\bk)$ is odd in momentum. If cleaner samples become available and display a significantly enhanced $T_c$, then the inter-SL scenario would remain a good candidate. Furthermore, the presence and location of nodes in the superconducting gap can give us important information since the inter-SL involves odd-parity states which necessarily display line nodes in the superconducting gap at $k_z=0$.


In this work, we do not discuss the pairing mechanism. For a realistic discussion, details of the FS in the heavy fermion regime are needed in order to investigate possible spin-fluctuation mechanisms. Also, the clarification of the nature of the hidden order observed at 0.4K and its association with multipolar order brings an interesting possibility for exotic pairing from multipolar interactions \cite{Sakai:2012,Tsujimoto:2014,Kotegawa:2003}. Moreover, the presence of quantum critical behavior evidenced by the temperature dependence of the specific heat at low temperatures suggests a scenario similar to $\beta$-YbAlB$_4$, for which a careful description of the crystal electric field states was key to understand its phenomenology \cite{Nakatsuji:2008,Matsumoto:2011,Ramires:2012,Ramires:2014}.

In summary, we have analyzed all possible superconducting order parameters for CeRh$_2$As$_2$ within the Ce-sites perspective. We find temperature versus magnetic field phase diagrams in qualitative agreement with experiments for both SOC and ILH dominated normal states. We have identified two possible scenarios for the two superconducting phases observed in CeRh$_2$As$_2$:  
i) even-odd intra-SL, associated with the previously proposed PDW scenario, and ii) new 
odd-odd inter-SL scenario. Our work stimulates more efforts in the field of complex locally noncentrosymmetric materials with extra internal degrees of freedom. 
We have preliminary evidence that the joint action of SOC and magnetic field might lead parity mixing in both low and high field phases.
We detail this possibility in a future more extended manuscript.
Further theoretical work analyzing the interplay of normal state parameters and magnetic field, as well as topological aspects are interesting directions for future work.

\begin{acknowledgments}
The authors acknowledge E. Hassinger and Y.-P. Huang for useful discussions.
\end{acknowledgments}

\bibliographystyle{apsrev4-1}
\bibliography{CeRh2As2}{}

\end{document}